\documentclass[a4paper,12pt]{article}
\usepackage[english]{babel}
\usepackage{ae}
\usepackage{amsmath,amssymb,amsfonts,euscript,amscd}
\usepackage{amsthm}
\usepackage{exscale,bm}
\usepackage[T1]{fontenc}
\usepackage[cp1251]{inputenc}
\usepackage[active]{srcltx}

\title{On 't Hooft--Polyakov Monopole, Julia--Zee Dyon, and Higgs Field, throughout the Generalized Bogomoln'yi Equations}
\author{\textbf{Lukasz Andrzej Glinka}\footnote{E-mail to: laglinka@gmail.com, lukaszglinka@wp.eu}}

\date{\empty}

\begin{document}

\maketitle
\thispagestyle{empty}
\begin{abstract}
In this paper, making use of the 't Hooft--Polyakov--Julia--Zee ansatz for the $SU(2)$ Yang--Mills--Higgs gauge field theory, we present the straightforward generalization of the Bogomoln'yi equations and its several consequences. Particularly, this is shown that this idea is able to generate new types of non-abelian both dyons and magnetic monopoles and, moreover, that within the new model the scalar field can be described through the Coulomb potential, whereas, up to a constant, the non-abelian gauge field becomes the Wu--Yang monopole.\\

\noindent \textbf{Keywords:} gauge field theories, SU(2), Yang--Mills--Higgs equations, 't Hooft--Polyakov monopole, Julia--Zee dyon, Bogomoln'yi equations, BPS limit, non-abelian dyons, non-abelian magnetic monopoles, Higgs field, Coulomb potential, Wu--Yang monopole
\end{abstract}
\newpage
\section{Yang--Mills--Higgs Equations}

The Yang--Mills--Higgs theory is the non-abelian Yang--Mills theory of the gauge field ${A}_\nu^a$ \cite{yangmills} coupled to the Higgs field $\varphi^a$ \cite{higgs}. For the $SU(N)$ gauge group, the Lagrangian has the form
\begin{equation}\label{lagymh}
  \mathcal{L}_{YMH}=-\dfrac{1}{4\mu_0}F^a_{\mu\nu}F_a^{\mu\nu}+\dfrac{\hslash c}{2}D_\mu\varphi^aD^\mu\varphi_a-\dfrac{\lambda\hslash c}{4}\left(\varphi^a\varphi_a-\varphi_0^2\right)^2,
\end{equation}
where $a=1,\ldots,N^2-1$ is a gauge group index, and
\begin{equation}
F^a_{\mu\nu}=\partial_\mu A_\nu^a-\partial_\nu A_\mu^a+gf^a_{bc} A^b_\mu A^c_\nu,\quad D_\mu=\partial_\mu-igT_b A^b_\mu,
\end{equation}
are the strength tensor and the gauge-covariant derivative, respectively, while $g$ is a coupling constant, and $f^a_{bc}$ are the (antisymmetric) structure constants of the Lie algebra $\mathfrak{su}(N)$ whose elements are the infinitesimal generators $T_a$ of $SU(N)$
\begin{equation}
  [T_a,T_b]=i{f_{ab}}^cT_c,\quad \textrm{tr}T_a=0, \quad T_a^\dag=T_a, \label{liecom}
\end{equation}
and $SU(N)$ is produced by formal exponentiation of $\mathfrak{su}(N)$. Furthermore,
\begin{equation}
\langle0|\varphi^a\varphi_a|0\rangle=\varphi_0^2=\dfrac{m^2c^2}{2\lambda\hslash^2},
\end{equation}
$m$ is a mass parameter, and $\lambda$ is a dimensionless constant. The energy density
\begin{equation}\label{hymh}
 \mathcal{H}_{YMH}=\dfrac{\varepsilon_0}{2}E^a_kE_a^k+\dfrac{1}{2\mu_0}B^a_kB_a^k+\dfrac{\hslash c}{2}D_k\varphi^aD^k\varphi_a+\dfrac{\lambda\hslash c}{4}\left(\varphi^a\varphi_a-\varphi_0^2\right)^2,
\end{equation}
where $E_a^k= cF_a^{k0}$ is the electric field and $B_a^k=\dfrac{1}{2}\varepsilon^{kij}F^a_{ij}$ is the magnetic field induction, is minimized $\mathcal{H}_{YMH}=0$ for the Higgs vacuum
\begin{eqnarray}
 \varphi^a\varphi_a=\varphi_0^2,\quad D_\mu\varphi^a=0,\quad F^a_{\mu\nu}=0,
\end{eqnarray}
which is degenerated, that is the set of vacuum expectation values of the Higgs field forms a sphere $S^2_\infty$ on which all the points are equivalent because the gauge transformation connects them. Moreover, for $\varphi_0\neq0$ there is the spontaneous symmetry breakdown mechanism $SU(N)\rightarrow U(1)^{N-1}$. The field equations for the Yang--Mills--Higgs theory (\ref{lagymh}) are
\begin{eqnarray}
D^\mu{D}_\mu\varphi_a+\lambda\left(\varphi^c\varphi_c-\varphi_0^2\right)\varphi_a&=&0,\label{ymh1}\\
\partial_\mu{F}_a^{\mu\nu}+gf_{abc} A^b_\mu{}F_c^{\mu\nu}&=&\mu_0\hslash cgf_{abc}\varphi^bD^\nu\varphi^c,\label{ymh2}
\end{eqnarray}
and the Bianchi identities
\begin{equation}
  \partial_\mu G_a^{\mu\nu}+gf_{abc} A^b_\mu G_c^{\mu\nu}=0.\label{ymh3}
\end{equation}
are satisfied for the dual field strength tensor $G_a^{\mu\nu}=\dfrac{1}{2}\varepsilon^{\mu\nu\kappa\lambda}F^a_{\kappa\lambda}$.

\section{'t Hooft--Polyakov--Julia--Zee Ansatz}

Dyon, a hypothetical object carrying a non-zero both electric and magnetic charges, was for the first time considered through Julian Seymour Schwinger \cite{sdyons}, who applied this model as the phenomenological alternative to quarks to describe a particle analogous to the flavor-neutral $\mathrm{J/\psi}$ meson consisting of a charm quark and a charm antiquark, known as charmonium, discovered soon later \cite{ting}. Particularly, the Yang--Mills--Higgs system was considered for the case of the $SU(2)$ gauge group, that is when the structure constants are the three-dimensional Levi--Civita symbols $\varepsilon_{abc}$ and the infinitesimal generators in the fundamental representation are given by the Pauli matrices $\sigma_a$ as $T_a=\dfrac{1}{2}\sigma_a$. In such gauge field theory, a smooth nonsingular solution to the Yang--Mills--Higgs equations, arising from the assumptions of spherical symmetry and time independence of both the non-abelian gauge field and the Higgs field
\begin{eqnarray}
 A_k^a=\dfrac{1-K\left(\xi\right)}{gr}\varepsilon^{a}_{ik}\dfrac{r^k}{r},\quad \varphi^a=\varphi_0\dfrac{H(\xi)}{\xi}\dfrac{r^a}{r},\quad \xi=\dfrac{\hslash}{e}g\varphi_0r,
\end{eqnarray}
 known as the 't Hooft--Polyakov ansatz, was considered in the pioneering papers authored through Gerardus 't Hooft \cite{thooft} and A.M. Polyakov \cite{polyakov}, which with the boundary conditions
 \begin{equation}
  K(\xi\rightarrow0)\rightarrow 1,\quad H(\xi\rightarrow0)<O(\xi),\quad  K(\xi\rightarrow\infty)\rightarrow0,\quad H(\xi\rightarrow\infty)\sim\xi,
 \end{equation}
 is known as the 't Hooft--Polyakov monopole. In the context of gauge field theories, dyons were for the first time considered as the solutions to the Yang--Mills--Higgs equations through Bernard Julia and Anthony Zee \cite{julia}, who supplemented the 't Hooft--Polyakov ansatz by
\begin{equation}
 A_0^a=\dfrac{J(\xi)}{gr}\dfrac{r^a}{r},\quad J(r\rightarrow0)\rightarrow0,\quad J(r\rightarrow\infty)\rightarrow Cr,
\end{equation}
where $C$ is a constant. The 't Hooft--Polyakov--Julia--Zee ansatz is compatible with the Lorentz gauge $\partial^\mu A^a_\mu=0$, which gives
\begin{equation}
\varepsilon^a_{ki}r^k\nabla^iK=0,
\end{equation}
that is $\vec{r}\times\vec{\nabla}K=0$, and, particularly, is satisfied through each spherically symmetric function $K$.

\section{Bogomol'nyi--Prasad--Sommerfield Limit}

The only known analytical solution occurs for the Bogomol'nyi--Prasad--Sommerfield (BPS) limit \cite{ps,bog}, that is $\lambda\rightarrow0$. Then, the Higgs field is massless but the full gauge invariance remains broken due to its non-zero vacuum expectation value. It is easy to see that the energy calculated by (\ref{hymh})
\begin{eqnarray}\label{hymhy1}
\!\!\!\!\!\!\!\!\!\!\!\!&&\mathcal{E}_{YMH}=\int d^3x V+\sqrt{\dfrac{\hslash c}{\mu_0}}\left(\dfrac{1}{c}\sin\alpha_c\int d^3x E_a^kD_k\varphi^a+\cos\alpha_c\int d^3x B_a^kD_k\varphi^a\right)\nonumber\\
\!\!\!\!\!\!\!\!\!\!\!\!&&+\dfrac{\varepsilon_0}{2}\int d^3x\left(E_a^k-c\sqrt{\mu_0\hslash c}\sin\alpha_c D^k\varphi_a\right)\left(E^a_k-c\sqrt{\mu_0\hslash c}\sin\alpha_c D_k\varphi^a\right)\nonumber\\
\!\!\!\!\!\!\!\!\!\!\!\!&&+\dfrac{1}{2\mu_0}\int d^3x\left(B_a^k-\sqrt{\mu_0\hslash c}\cos\alpha_c D^k\varphi_a\right)\left(B^a_k-\sqrt{\mu_0\hslash c}\cos\alpha_c D_k\varphi^a\right),
\end{eqnarray}
where $V=\dfrac{\lambda\hslash c}{4}\left(\varphi^a\varphi_a-\varphi_0^2\right)^2$ is the scalar field potential., and
\begin{equation}
  \tan\alpha_c=\dfrac{qc}{q_M}.
\end{equation}
The minimum of (\ref{hymhy1}) in the BPS limit occurs for the Bogomol'nyi equations
\begin{eqnarray}
  E_a^k&=&c\sqrt{\mu_0\hslash c}\sin\alpha_c D^k\varphi_a,\label{BOG2}\\
  B_a^k&=&\sqrt{\mu_0\hslash c}\cos\alpha_c D^k\varphi_a.\label{BOG1}
\end{eqnarray}
The Gauss law for electricity for $\mathbf{E}=\mathbf{E}^a\dfrac{\varphi_a}{\varphi_0}$ gives
\begin{equation}\label{gausel}
\dfrac{1}{\varphi_0}\int d^3x E_a^kD_k\varphi^a=\int\mathbf{E}d\mathbf{S}=\dfrac{q}{\varepsilon_0},
\end{equation}
whereas the Gauss law for magnetism for $\mathbf{B}=\mathbf{B}^a\dfrac{\varphi_a}{\varphi_0}$ leads to
\begin{equation}\label{gausmag}
\dfrac{1}{\varphi_0}\int d^3x B_a^kD_k\varphi^a=\int\mathbf{B}d\mathbf{S}=\mu_0q_M,
\end{equation}
and, in this manner,
\begin{equation}
  q=\dfrac{4\pi\varepsilon_0c}{g}C=\dfrac{q_M}{c}C,\quad C=\tan\alpha_c.
\end{equation}
Consequently, one receives
\begin{equation}\label{hymhy1a}
 \mathcal{E}_{YMH}=\hslash c\varphi_0\sqrt{4\pi \alpha}\left(\dfrac{q}{e}\sin\alpha_c+\dfrac{q_M}{ec}\cos\alpha_c\right)
 +\dfrac{\lambda\hslash c}{4}\int dV\left(\varphi^a\varphi_a-\varphi_0^2\right)^2,
\end{equation}
where $\alpha=\dfrac{e^2}{4\pi\varepsilon_0\hslash c}$ is the fine-structure constant. Through Einstein's relation $\mathcal{E}_{YMH}=Mc^2$, the effective mass $M$ satisfies the Bogomol'nyi bound
\begin{equation}
  M\geqslant \dfrac{\hslash\varphi_0}{c}\sqrt{4\pi\alpha}\left(\dfrac{q}{e}\sin\alpha_c+\dfrac{q_M}{ec}\cos\alpha_c\right),
\end{equation}
and the equality holds in the BPS limit. For the 't Hooft--Polyakov monopole, in the BPS limit, one has
\begin{equation}
  M=\varphi_0\sqrt{4\pi\alpha}\dfrac{q_M\hslash}{ec^2},
\end{equation}
the total energy density of such a solution is
\begin{equation}
  \mathcal{H}=\hslash cD_k\varphi^aD^k\varphi_a=\dfrac{1}{\cos^2\alpha_c}\dfrac{\mathbf{B}^2}{\mu_0}=\dfrac{\varepsilon_0}{\sin^2\alpha_c}\mathbf{E}^2,
\end{equation}
and the Bogomol'nyi equations (\ref{BOG1})-(\ref{BOG2}) become
\begin{eqnarray}
E^a_k=0,\quad B_a^k=\sqrt{\mu_0\hslash c}D^k\varphi_a.
\end{eqnarray}
The subject of non-abelian both magnetic monopoles and dyons has already met a certain interest, Cf. the Refs. \cite{shnir} and references therein.

\section{Generalization of Bogomol'nyi Equations}

Let us generalize the BPS solution to any $\lambda$. Considering the Yang--Mills--Higgs energy in the form $\mathcal{E}_{YMH}=\mathcal{E}_{YM}+\mathcal{E}_{H}$, where
\begin{eqnarray}\label{hymhym}
\mathcal{E}_{YM}=\int d^3x \left(\dfrac{\varepsilon_0}{2}\tilde{E}^a_k\tilde{E}_a^k+\dfrac{1}{2\mu_0}\tilde{B}^a_k\tilde{B}_a^k\right),
\end{eqnarray}
where
\begin{eqnarray}
\tilde{E}^a_k&=&E^a_k-c\sqrt{\mu_0\hslash c}\alpha_1D_k\varphi^a-c\sqrt{\mu_0V}\beta_1\tilde{\delta}_k^a,\quad \tilde{E}_a^k=\delta_{ab}\delta^{kj}\tilde{E}^b_j\\
\tilde{B}^a_k&=&B^a_k-\sqrt{\mu_0\hslash c}\alpha_2D_k\varphi^a-\sqrt{\mu_0V}\beta_2\tilde{\delta}'^a_k,\quad \tilde{B}_a^k=\delta_{ab}\delta^{kj}\tilde{B}^b_j,
\end{eqnarray}
and
\begin{eqnarray}
\tilde{\delta}^a_k&=&\delta_1\delta^a_k+\delta_2\dfrac{\varphi^a\varphi_k}{|\varphi|^2},\quad\tilde{\delta}^k_a=\delta_{ab}\delta^{kj}\tilde{\delta}^b_j,\\ \tilde{\delta}'^a_k&=&\delta_1'\delta^a_k+\delta_2'\dfrac{\varphi^a\varphi_k}{|\varphi|^2},\quad\tilde{\delta}'^k_a=\delta_{ab}\delta^{kj}\tilde{\delta}'^b_j,
\end{eqnarray}
 and $\alpha_1$, $\alpha_2$, $\beta_1$, $\beta_2$, $\delta_1$, $\delta_2$, $\delta_1'$, $\delta_2'$ are dimensionless constants, and
\begin{eqnarray}
\!\!\!\!\!\!\!\!\!\!\!\!\!\!
&&\mathcal{E}_{H}=\sqrt{\dfrac{\hslash c}{\mu_0}}\int d^3x\left(\alpha_1\dfrac{E^a_k}{c}+\alpha_2B^a_k\right)D^k\varphi_a+\label{enedue}\\
\!\!\!\!\!\!\!\!\!\!\!\!\!\!
&&\int d^3x\left(\beta_1\tilde{\delta}^k_a\dfrac{E^a_k}{c}+\beta_2\tilde{\delta}'^k_aB^a_k\right)\sqrt{\dfrac{V}{\mu_0}}-\int d^3x\left(\alpha_1\beta_1\tilde{\delta}^k_a+\alpha_2\beta_2\tilde{\delta}'^k_a\right)\sqrt{\hslash cV}D_k\varphi^a,\nonumber
\end{eqnarray}
where the following consistency conditions hold
 \begin{eqnarray}
  &&\alpha_1^2+\alpha_2^2=1,\label{CON1}\\
  &&\dfrac{\beta_1^2}{2}\tilde{\delta}^k_a\tilde{\delta}_k^a+\dfrac{\beta_2^2}{2}\tilde{\delta}'^k_a\tilde{\delta}'^a_k=1.\label{CON2}
\end{eqnarray}
Applying the Gauss laws (\ref{gausel}) and (\ref{gausmag}), one receives
\begin{equation}
\sqrt{\dfrac{\hslash c}{\mu_0}}\int d^3x\left(\alpha_1\dfrac{E^a_k}{c}+\alpha_2B^a_k\right)D^k\varphi_a=
\hslash c\varphi_0\sqrt{4\pi\alpha}\left(\alpha_1\dfrac{q}{e}+\alpha_2\dfrac{q_M}{ec}\right),
\end{equation}
and after careful calculations, one receives
\begin{eqnarray}
\mathcal{E}_{YMH}&=&\mathcal{E}_{YM}+\hslash c\varphi_0\sqrt{4\pi\alpha}\left(\alpha_1\dfrac{q}{e}+\alpha_2\dfrac{q_M}{ec}\right)+\int d^3xV\label{enedue1}\\&+&\int
d^3x\left(\dfrac{\beta_1}{c}\tilde{E}^a_k\tilde{\delta}^k_a+\beta_2\tilde{B}^a_k\tilde{\delta}'^k_a\right)\sqrt{\dfrac{V}{\mu_0}}.\nonumber
\end{eqnarray}
The minimum of (\ref{enedue1}) is not only the Higgs vacuum, but also $\tilde{E}^a_k=0$ and $\tilde{B}^a_k=0$, giving the BPS limit, or equivalently
\begin{eqnarray}
E^a_k&=&c\sqrt{\mu_0}\left(\sqrt{\hslash c}\alpha_1D_k\varphi^a+\sqrt{V}\beta_1\tilde{\delta}_k^a\right),\label{GLI1}\\
B^a_k&=&\sqrt{\mu_0}\left(\sqrt{\hslash c}\alpha_2D_k\varphi^a+\sqrt{V}\beta_2\tilde{\delta}'^a_k\right),\label{GLI2}
\end{eqnarray}
and then the effective mass is
\begin{equation}
M=\dfrac{\hslash\varphi_0}{c}\sqrt{4\pi\alpha}\left(\alpha_1\dfrac{q}{e}+\alpha_2\dfrac{q_M}{ec}\right)+\dfrac{1}{c^2}\int d^3xV.\label{masss}
\end{equation}
If, moreover, this mass identically vanishes, that is
\begin{equation}
\int d^3xV=-\hslash c\varphi_0\sqrt{4\pi\alpha}\left(\alpha_1\dfrac{q}{e}+\alpha_2\dfrac{q_M}{ec}\right),
\end{equation}
and
\begin{equation}
  \dfrac{\alpha_1}{\alpha_2}=-\dfrac{q_M}{qc},
\end{equation}
then one has the condition
\begin{equation}
\int d^3xV=0,\label{volpot}
\end{equation}
which includes, as the particular situation, the BPS limit, but in general $V$ is not necessary zero. Similarly, for (\ref{masss}) the particular case
\begin{equation}
  M=\dfrac{\hslash \varphi_0}{c}\sqrt{4\pi\alpha}\left(\alpha_1\dfrac{q}{e}+\alpha_2\dfrac{q_M}{ec}\right),
\end{equation}
is true if and only if (\ref{volpot}), also for the BPS limit. In general, the bound
\begin{equation}
  M\geqslant \dfrac{\hslash\varphi_0}{c}\sqrt{4\pi\alpha}\left(\alpha_1\dfrac{q}{e}+\alpha_2\dfrac{q_M}{ec}\right),
\end{equation}
which for $\alpha_1=\sin\alpha_c$, $\alpha_2=\cos\alpha_c$ is the Bogomol'nyi bound, is true for
\begin{equation}
\int d^3xV\geqslant 0.\label{volpot1}
\end{equation}

The condition (\ref{CON1}) gives eight inequivalent solutions
\begin{eqnarray}
 \alpha_1=\pm\sin\alpha_c&\& &\alpha_2=\pm\cos\alpha_c,\\
 \alpha_1=\pm\cos\alpha_c&\&&\alpha_2=\pm\sin\alpha_c,
\end{eqnarray}
whereas the condition (\ref{CON2}) presented in the following form
\begin{eqnarray}
\dfrac{\beta_1^2}{2}\left(D\delta_1^2+\delta_2^2+2\delta_1\delta_2\right)+\dfrac{\beta_2^2}{2}\left(D\delta_1'^2+\delta_2'^2+2\delta_1'\delta_2'\right)=1,\quad D=\delta^a_k\delta_a^k,\label{CON2a}
\end{eqnarray}
and parameterized as $A^2+B^2=1$, also leads to eight solutions
\begin{eqnarray}
 A=\pm\sin\beta_c&\& &B=\pm\cos\beta_c,\\
 A=\pm\cos\beta_c&\& &B=\pm\sin\beta_c,
\end{eqnarray}
where $\beta_c$ must not be related to $\alpha_c$ in general. Then, one obtains
\begin{eqnarray}
  \delta_2=\delta_1\left(-1\pm\sqrt{\dfrac{2A^2}{\beta_1^2\delta_1^2}+1-D}\right),\quad
  \delta_2'=\delta_1'\left(-1\pm\sqrt{\dfrac{2B^2}{\beta_2^2\delta_1'^2}+1-D}\right),\\
  \delta_1=\dfrac{\delta_2}{D}\left(-1\pm\sqrt{\dfrac{2A^2}{\beta_1^2\delta_2^2}+1-D}\right),\quad
  \delta_1'=\dfrac{\delta_2'}{D}\left(-1\pm\sqrt{\dfrac{2B^2}{\beta_2^2\delta_2'^2}+1-D}\right),
\end{eqnarray}
what is consistent for
\begin{eqnarray}
  \beta_1^2\delta_1^2\geqslant\dfrac{2A^2}{D-1},\quad
  \beta_2^2\delta_1'^2\geqslant\dfrac{2B^2}{D-1},\quad
  \beta_1^2\delta_2^2\geqslant\dfrac{2A^2}{D-1},\quad
  \beta_2^2\delta_2'^2\geqslant\dfrac{2B^2}{D-1}.
\end{eqnarray}

The 't Hooft--Polyakov--Julia--Zee ansatz
\begin{equation}\label{hpjz}
 \!A_\mu^a=\dfrac{\hslash\varphi_0}{e}\left[\dfrac{J(\xi)}{\xi}\dfrac{r^a}{r},\dfrac{1-K(\xi)}{\xi}\dfrac{\varepsilon^a_{ik}r^k}{r}\right],~~ \varphi^a=\varphi_0\dfrac{H(\xi)}{\xi}\dfrac{r^a}{r},~~ \xi=\dfrac{\hslash}{e}g\varphi_0r,
\end{equation}
leads to the relations
\begin{eqnarray}
  E^a_k&=&c\dfrac{\hslash^2\varphi^2_0}{e^2}\dfrac{g}{\xi^2}\left[\left(\xi\dfrac{dJ}{d\xi}-J(K+1)\right)\dfrac{r^ar_n}{r^2}+JK\delta^a_n\right],\label{gyt0a}\\
  B^a_k&=&\dfrac{\hslash^2\varphi^2_0}{e^2}\dfrac{g}{\xi^2}\left[\left(1-K^2+\xi\dfrac{dK}{d\xi}\right)\dfrac{r^ar_n}{r^2}-\xi\dfrac{dK}{d\xi}\delta^a_n\right],\label{gyt1a}\\
  D_i\varphi^a&=&\dfrac{\hslash\varphi^2_0}{e}\dfrac{g}{\xi^2}\left[\left(\xi\dfrac{dH}{d\xi}-KH-H\right)\dfrac{r^ar_i}{r^2}+\delta^a_iKH\right],\label{gyt2a}
\end{eqnarray}
which, along with (\ref{GLI1}) and (\ref{GLI2}), lead to the system of equations for a dyon
\begin{eqnarray}
\!\!\!\!\!\!\!\!\!\!\xi\dfrac{dJ}{d\xi}-J(K+1)&=&\alpha_1\sqrt{4\pi\alpha}\left(\xi\dfrac{dH}{d\xi}-H(K+1)+\dfrac{\beta_1\delta_2e}{\alpha_1\hslash g\varphi_0^2}\sqrt{\dfrac{V}{\hslash c}}\xi^2\right),\label{GLIN1}\\
\!\!\!\!\!\!\!\!\!\!JK&=&-\alpha_1\sqrt{4\pi\alpha}\left(KH+\dfrac{\beta_1\delta_1e}{\alpha_1\hslash g\varphi_0^2}\sqrt{\dfrac{V}{\hslash c}}\xi^2\right),\label{GLIN2}\\
\!\!\!\!\!\!\!\!\!\!\xi\dfrac{dK}{d\xi}-K^2+1&=&\alpha_2\sqrt{4\pi\alpha}\left(\xi\dfrac{dH}{d\xi}-H(K+1)+\dfrac{\beta_2\delta_2'e}{\alpha_2\hslash g\varphi_0^2}\sqrt{\dfrac{V}{\hslash c}}\xi^2\right),\label{GLIN3}\\
\!\!\!\!\!\!\!\!\!\!\xi\dfrac{dK}{d\xi}&=&-\alpha_2\sqrt{4\pi\alpha}\left(KH+\dfrac{\beta_2\delta_1'e}{\alpha_2\hslash g\varphi_0^2}\sqrt{\dfrac{V}{\hslash c}}\xi^2\right).\label{GLIN4}
\end{eqnarray}

For the 't Hooft --Polyakov monopole, $J=0$, (\ref{GLIN1}), (\ref{GLIN2}), (\ref{GLIN3}) and (\ref{GLIN4}) are
\begin{eqnarray}
\xi\dfrac{dH}{d\xi}-H(K+1)&=&-\dfrac{e}{\hslash g\varphi_0^2}\dfrac{\beta_1\delta_2}{\alpha_1}\sqrt{\dfrac{V}{\hslash c}}\xi^2,\label{GLIN1a}\\
KH&=&-\dfrac{e}{\hslash g\varphi_0^2}\dfrac{\beta_1\delta_1}{\alpha_1}\sqrt{\dfrac{V}{\hslash c}}\xi^2,\label{GLIN2a}\\
\xi\dfrac{dK}{d\xi}-K^2+1&=&\alpha_2\sqrt{4\pi\alpha}\dfrac{e}{\hslash g\varphi_0^2}\left(\dfrac{\beta_2\delta_2'}{\alpha_2}-\dfrac{\beta_1\delta_2}{\alpha_1}\right)\sqrt{\dfrac{V}{\hslash c}}\xi^2,\label{GLIN3a}\\
\xi\dfrac{dK}{d\xi}&=&-\alpha_2\sqrt{4\pi\alpha}\dfrac{e}{\hslash g\varphi_0^2}\left(\dfrac{\beta_2\delta_1'}{\alpha_2}-\dfrac{\beta_1\delta_1}{\alpha_1}\right)\sqrt{\dfrac{V}{\hslash c}}\xi^2,\label{GLIN4a}
\end{eqnarray}
and, therefore, one obtains the solutions
\begin{eqnarray}
K&=&\pm\sqrt{1+\dfrac{G}{\varphi_0^2}\sqrt{\dfrac{V}{\hslash c}}\xi^2}\label{GLIN5},\\
H&=&\mp\dfrac{e}{\hslash g\varphi_0^2}\dfrac{\beta_1\delta_1}{\alpha_1}\dfrac{\sqrt{\dfrac{V}{\hslash c}}\xi^2}{\sqrt{1+\dfrac{G}{\varphi_0^2}\sqrt{\dfrac{V}{\hslash c}}\xi^2}}.\label{GLIN6}
\end{eqnarray}
Consequently, the Higgs field and the gauge field are
\begin{eqnarray}
\varphi^a&=&\dfrac{\mp\dfrac{\beta_1\delta_1}{\alpha_1}\sqrt{\dfrac{V}{\hslash c}}}{\sqrt{1+\dfrac{G}{\varphi_0^2}\sqrt{\dfrac{V}{\hslash c}}\xi^2}}r^a,\\
 A_\mu^a&=&\left[0,\left(1\mp\sqrt{1+\dfrac{G}{\varphi_0^2}\sqrt{\dfrac{V}{\hslash c}}\xi^2}\right)\varepsilon^a_{ik}\dfrac{r^k}{gr^2}\right],\label{eurasia}
\end{eqnarray}
whereas simultaneously the scalar field potential $V$ satisfies the following differential equation
\begin{equation}\label{scfieq}
\dfrac{dV}{d\xi}=\dfrac{-4\pm4\dfrac{\delta_2+\delta_1}{\delta_1}\left(1+\dfrac{G}{\varphi_0^2}\sqrt{\dfrac{V}{\hslash c}}\xi^2\right)^{3/2}}{2+\dfrac{G}{\varphi_0^2}\sqrt{\dfrac{V}{\hslash c}}\xi^2}\dfrac{V}{\xi},
\end{equation}
where we have introduced the constant
\begin{equation}
  G=\dfrac{e}{\hslash g}\sqrt{4\pi\alpha}\left(\dfrac{\alpha_2}{\alpha_1}\beta_1\left(\delta_2+\delta_1\right)-\beta_2\left(\delta_1'+\delta_2'\right)\right).
\end{equation}

Interestingly, in the BPS limit, the 't Hooft--Polyakov monopole is described through $H=0$ and $K=\pm1$. Then, $\varphi^a=0$, for $K=1$ also $A^a_\mu=0$, while for $K=-1$
\begin{equation}\label{wy}
 A^a_\mu=2\left(A^a_{\mu\nu}\right)_{WY},\quad \left(A^a_{\mu\nu}\right)_{WY}=\left[0,\varepsilon^a_{ik}\dfrac{r^k}{gr^2}\right],
\end{equation}
where $\left(A^a_{\mu\nu}\right)_{WY}$ is the Wu--Yang monopole solving the $SU(2)$ Yang--Mills theory without the Higgs field \cite{wy}. Nevertheless, one can also consider the situation wherein $G=0$, or, equivalently
\begin{equation}\label{G0}
\dfrac{\beta_1}{\alpha_1}\left(\delta_2+\delta_1\right)=\dfrac{\beta_2}{\alpha_2}\left(\delta_1'+\delta_2'\right).
\end{equation}
Then, the gauge field is either trivial or (\ref{wy}), while the Higgs field is
\begin{equation}
  \varphi^a=\mp\dfrac{\beta_1\delta_1}{\alpha_1}\sqrt{\dfrac{V}{\hslash c}}r^a,
\end{equation}
and, moreover, the equation (\ref{scfieq}) becomes
\begin{equation}\label{scfieq1}
\dfrac{dV}{V}=2\left(-1\pm\dfrac{\delta_2+\delta_1}{\delta_1}\right)\dfrac{d\xi}{\xi},
\end{equation}
and its straightforward integration gives two possible solutions
\begin{eqnarray}
V_+(\xi)&=&V_+(\xi_0)\left|\dfrac{\xi}{\xi_0}\right|^{\delta_2/\delta_1},\\ V_-(\xi)&=&V_-(\xi_0)\left|\dfrac{\xi}{\xi_0}\right|^{-2-\delta_2/\delta_1},
\end{eqnarray}
where $\xi_0$ and $V(\xi_0)$ are the integration constants. However, for the special case $\delta_2=-\delta_1$ and $V_+(\xi_0)=V_-(\xi_0)=V_0$, one obtained the unique solution
\begin{equation}
V(\xi)=V_0\left|\dfrac{\xi}{\xi_0}\right|^{-1}\sim \dfrac{1}{r},
\end{equation}
which is the Coulomb potential. In such a situation, taking $\xi_0=\dfrac{\hslash g}{e}\varphi_0r_0$ and $V_0=\dfrac{\lambda\hslash c}{4}\dfrac{1}{r_0^4}$, where $r_0$ is any initial value of $r$, which particularly can be identified with a minimal scale, the configuration of fields is established throughout the gauge field given through the solution (\ref{wy}), whereas the Higgs field has the following form
\begin{equation}
  \varphi^a=\mp\dfrac{\beta_1\delta_1}{2\alpha_1}\sqrt{\dfrac{\lambda}{r_0r}}\dfrac{r^a}{r_0}.
\end{equation}

\section{Summary}
We have seen that generalization of the Bogomoln'yi equations is realizable, and leads to potentially new results for non-abelian both dyons and magnetic monopoles. It should be emphasized that the idea of this paper was based on the author's monograph \cite{glinka}, and we hope for its further development.

\end{document}